\documentclass[pdflatex,sn-mathphys-num]{sn-jnl}
\usepackage{graphicx}
\usepackage{multirow}
\usepackage{amsmath,amssymb,amsfonts}
\usepackage{amsthm}
\usepackage{mathrsfs}
\usepackage[title]{appendix}
\usepackage{xcolor}
\usepackage{textcomp}
\usepackage{manyfoot}
\usepackage{booktabs}
\usepackage{algorithm}
\usepackage{algorithmicx}
\usepackage{algpseudocode}
\usepackage{listings}
\usepackage{siunitx}
\usepackage{adjustbox}
\begin{document}
\title{Binary Classification of Signal and Background Triggers of a Transition Edge Sensor Using Convolutional Neural Networks}

\author[1,*]{Elmeri Rivasto}
\author[2]{Katharina-Sophie Isleif}
\author[3]{Friederike Januschek}
\author[3]{Axel Lindner}
\author[1]{Manuel Meyer}
\author[2]{Gulden Othman}
\author[2]{Jos\'e Alejandro Rubiera Gimeno}
\author[3]{Christina Schwemmbauer}

\affil[1]{CP3-origins, Department of Physics, Chemistry and Pharmacy, University of Southern Denmark, Campusvej 55, 5230 Odense, Denmark}
\affil[2]{Helmut-Schmidt-Universität (HSU), Holstenhofweg 85, 22043 Hamburg, Germany}
\affil[3]{Deutsches Elektronen-Synchrotron DESY, Notkestr. 85, 22607 Hamburg, Germany}
\affil[*]{Contact: rivasto@cp3.sdu.dk}

\abstract{\noindent The Any Light Particle Search~II (ALPS~II) is a light shining through a wall experiment probing the existence of axions and axion-like particles using a 1064\,nm laser source. While ALPS II is already taking data using a heterodyne based detection scheme, cryogenic transition edge sensor (TES) based single-photon detectors are planned to expand the detection system for cross-checking the potential signals, for which a sensitivity on the order of 10\textsuperscript{-24}\,W is required. In order to reach this goal, we have investigated the use of convolutional neural networks (CNN) as binary classifiers to distinguish the experimentally measured 1064\,nm photon triggered (light) pulses from background (dark) pulses. Despite rigorous hyperparameter optimization, the CNN based binary classifier did not outperform our previously optimized cut-based analysis in terms of detection significance. Our findings suggest that training confusion, introduced by near-1064\,nm black-body photon triggers in the extrinsics background, is a significant factor limiting the CNNs performance for the associated dataset. The fiber coupled black-body radiation was identified as the limiting background source as concluded in our previous works. Given our results, we recommend that future studies explore regression-based CNNs, placing greater emphasis on the use of standardized and carefully structured training data rather than on extensive hyperparameter optimization. While the presented results and associated conclusions are obtained for TES designed to be used in the ALPS~II experiment, they should hold equivalently well for any device whose output signal can be considered as a univariate time trace.}

\maketitle

\section{Introduction}
\noindent Transition edge sensors (TES) are superconducting microcalorimeters that are voltage-biased within the region of superconducting phase transition where the resistance of the TES changes steeply with temperature \cite{Irwin2005transition}. Here, the absorption of a single photon heats up the TES sufficiently to result in a significant change in its bias current. These current perturbations are detected by an iductively coupled superconducting quantum interference device (SQUID). Unlike many other single-photon detectors, such as superconducting nanowire single-photon detectors (SNSPDs), TESs are capable of measuring the energy of the absorbed photons over a wide range of wavelengths. Their energy resolution together with high quantum efficiency and microsecond-scale dead time \cite{Lita2008counting, Hattori2022optical, DeLucia2024transition} make TESs important tools widely used in quantum computing \cite{Lita2010superconducting, Karasik2011nanobolometers, Mattioli2015photon, Hummatov2023fast, Li2023multi}, space and astrophysics experiments \cite{Romani1999first, Bruijn2003development, Goldie2011ultra, Bergen2016design, Appel2022calibration} along with particle physics and dark matter searches \cite{Miyazaki2020dark, Angloher2024doubletes, Romani2024transition}. 

A TES is planned to be used in a future science run of the \textit{Any Light Particle Search II} (ALPS~II) at DESY Hamburg (Germany) \cite{Bahre2013any}. ALPS II aims to probe the existence of axions and axion-like particles (ALPs) \cite{Shah2021tes, Gimeno2023tes} and is essentially a \textit{light shining through a wall} experiment featuring a powerful 1064\,nm laser beam that is shone into a 106\,m long resonant \textit{production cavity} that is in 5.3\,T magnetic field. While the propagation of the light is blocked by an opaque wall, the theoretically proposed photon--axion oscillation enables 1064\,nm photons to emerge on the other side of the optical barrier in a \textit{regeneration cavity} (symmetrical to the production cavity) \cite{Sikivie1983experimental, Isleif2022any}. The detection of these reconverted photons requires an extremely sensitive detection scheme achievable with TESs \cite{Shah2021tes, Gimeno2023tes}. The target sensitivity for the conversion rate lies in the range of $10^{-5}$\,Hz (one photon per day) setting the upper limit for the background rate required for the statistically significant detection of axions and ALPs within the practically feasible 20~days measurement time \cite{Shah2021tes}. 

In the recent years machine learning (ML) methods have been recognized as useful tools in various fields of physics \cite{Carleo2019machine}. ML approaches have been widely implemented for various tasks associated with particle physics experiments, including high level data analysis, jet tagging and data simulation \cite{Dery2017weakly, Barnard2017parton, Collado2021learning, Du2021deep, Graczykowski2022using}. ML based analysis of time traces, in particular, has been widely implemented in nuclear spectroscopy \cite{Zehtabvar2024review}, where neural networks have demonstrated significant performance in noise suppression \cite{Regadio2019filtering}. Lately, ML methods have also been used to detect hints from axion-like particles in LHC \cite{Ren2021detecting} and pulsar dispersion measurements \cite{Shi2025machine}, and are planned to be implemented for analyzing data in the SuperCDMS direct dark matter experiment \cite{Cushman2024strategies}. Most relevant for our interest, Manenti \textit{et al.} \cite{Manenti2024dark} have recently applied unsupervised ML models to study the background of a TES system that is not connected to the outside of a cryostat by an optical fiber (\textit{intrinsic background}). They report that the majority of the observed background pulses originate from high-energy events associated with radioactive decays and secondary cosmic-ray particles. The resulting dark counts are easy to distinguish from low-energy photon triggers by simply comparing the pulse shapes, as already concluded in our previous work \cite{Meyer2024first}. This is because the energy released from the various high-energy events is likely to be deposited within the underlying substrate rather than the TES itself. Due to slow diffusion of heat from the substrate to the TES, the intrinsic dark counts have significantly larger rise and decay times when compared with typical photon induced pulses where the energy is deposited directly to the TES \cite{Manenti2024dark, RubieraGimeno2024optimizing}.

While the unsupervised ML models have been mainly used for qualitative categorization of the recorded pulses, supervised ML models are better suited for actual quantitative background suppression. One can expect the state-of-the-art supervised ML models to outperform the capabilities of traditional data processing techniques. We have successfully implemented this in the past for \textit{intrinsic background} \cite{Meyer2024first}. In this work, we expand on this study by also considering the \textit{extrinsic background} measured while an optical fiber links the lab environment to the TES inside the dilution refrigerator. This mainly introduces an additional background resulting from fiber coupled black-body radiation. The black-body photons have been identified as the limiting background for our experimental setup \cite{Gimeno2023tes, Gimeno2024tes, Gimeno2025simulation}, and have been previously addressed using a traditional cut-based analysis without relying on machine learning \cite{RubieraGimeno2024optimizing}. We want to point out that the black-body background rate is ultimately determined by the energy resolution of the TES since higher energy resolution (smaller quantitative value) enables more reliable distinction between the signal and the background photons. Thus, different analysis methods addressing noise suppression differently can have significant effects on the background rates \cite{Gimeno2025simulation}. For example, we have previously found that performing fits to the measured TES pulses in frequency domain instead of time domain results in 2-fold improvement in the energy resolution \cite{Gimeno2024tes, RubieraGimeno2024optimizing}.

\begin{figure*}[t!]
\begin{center}
\includegraphics[width=10.0cm]{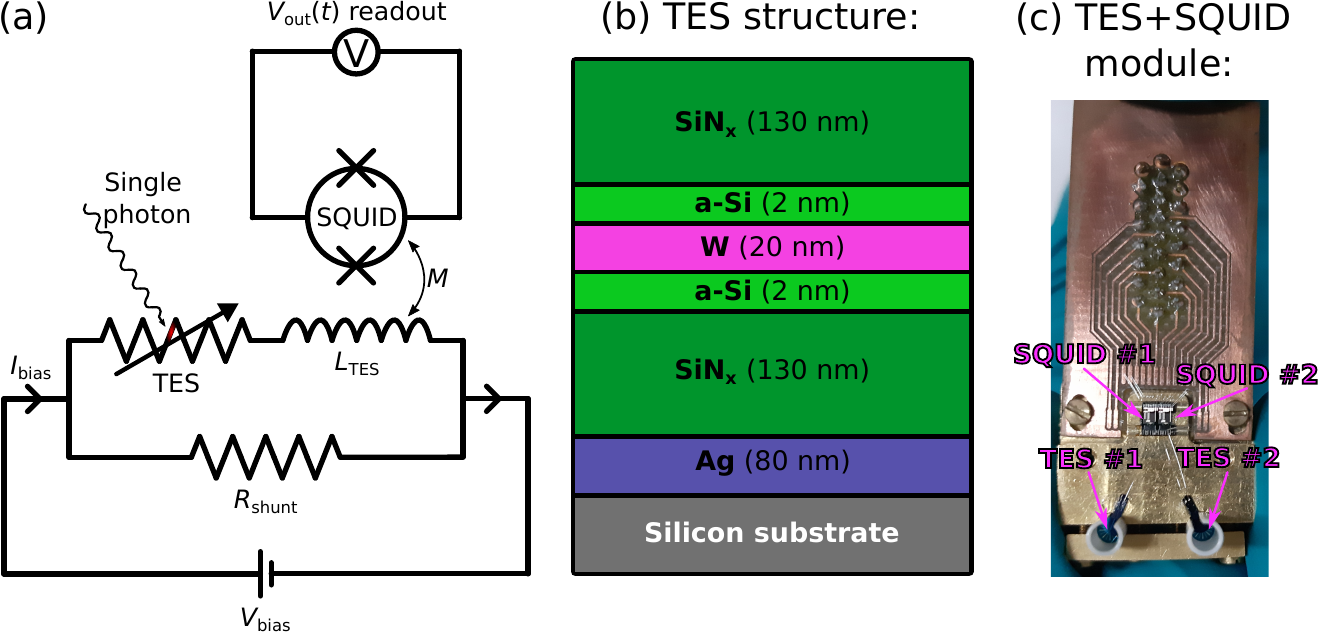}
\caption{(a) A circuit diagram of the experimental TES setup. Applying the constant bias voltage ($V_\mathrm{bias}$) and the shunt resistor ($R_\mathrm{shunt}$) parallel to the TES allows to set the TES working point (negative electrothermal feedback). (b) A schematic illustration of the layer structure of the TES optimized for detecting 1064\,nm photons, where the 20\,nm thick superconducting W layer with a surface area of $25\,\mathrm{\mu m}\times25\,\mathrm{\mu m}$ acts as the active material. (c) Picture of the used TES+SQUID module, including two TESs and their associated SQUIDs. }
\label{TES_schematics_Figure_ab}
\end{center}
\end{figure*}

\begin{figure*}[t!]
\begin{center}
\includegraphics[width=6.0cm]{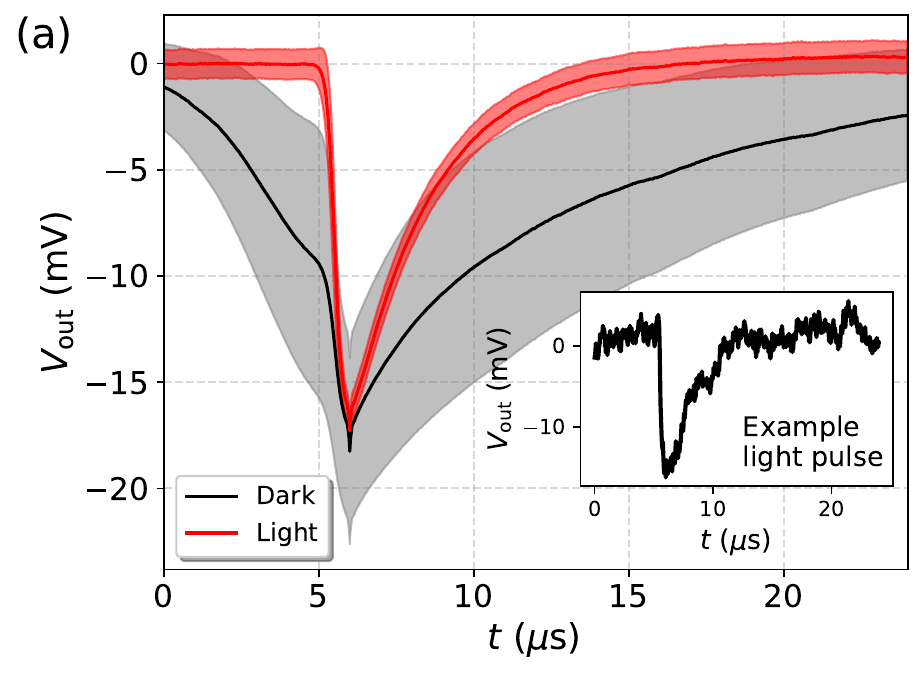}
\includegraphics[width=6.0cm]{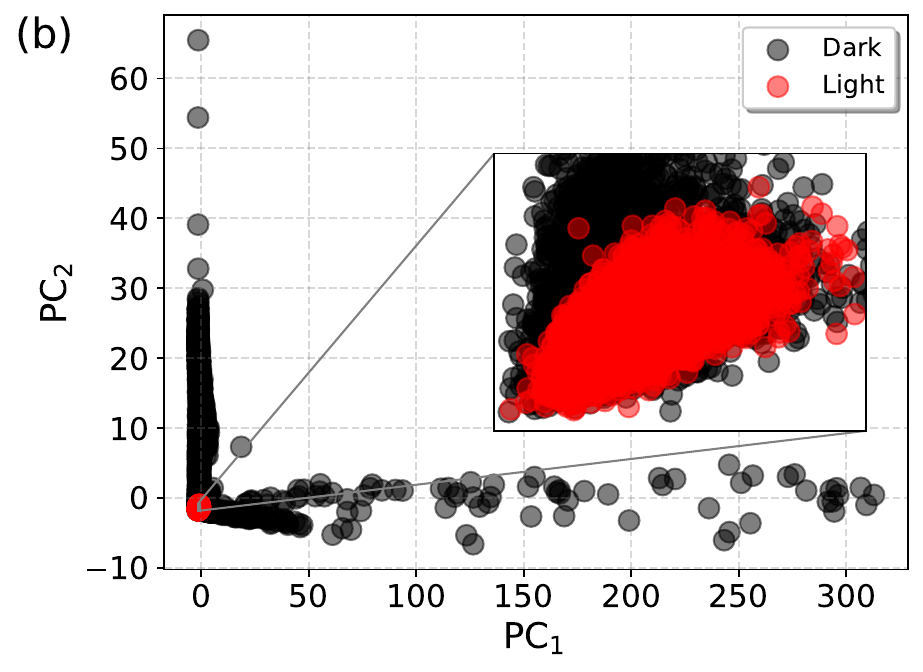}
\caption{(a) The average measured light pulse and dark pulses, where the shaded regions represent the associated standard deviations. The inset presents a randomly chosen light pulse as an example of the signal and noise shapes. (b) Principal Component Analysis (PCA) scatter plot showing the projection of pulse feature vectors ($\tau_\mathrm{rise}$, $\tau_\mathrm{decay}$, $\chi_\mathrm{ph}^2$, $\mathrm{V}_\mathrm{min,\,FFT}$, $\chi^2_\mathrm{FFT}$) onto the first two principal components (PC$_1$ and PC$_2$). The inset shows a close-up of the cluster associated with light pulses, showing overlap with some of the dark pulses measured in extrinsic background. }
\label{pulses_and_PCA_Figure_ab}
\end{center}
\end{figure*}

Here, for the first time, we are trying to further improve the rejection of near-1064\,nm black-body photons using convolutional neural networks (CNN) that are considered as the state-of-the-art machine learning model for univariate time-series classification. Ultimately, the goal is to reach a background rate below $10^{-5}\,\mathrm{Hz}$ while maintaining a tolerable rate for correctly classified signal pulses (analysis efficiency). The CNNs expand the architecture of the conventional multi-layer perceptrons (feedforward neural networks) via the introduction of convolutional layers that apply different filters (kernels) to the input data enabling the efficient extraction of spatial (or temporal) patterns \cite{Li2021survey}. The CNNs remain the underlying technology behind the state-of-the-art image classification ML models \cite{Lecun2015deep, Rawat2017deep, Li2021survey}, also covering the univariate time series classifiers \cite{Foumani2023deep, Wang2017time, Ismail2020inceptiontime, Dempster2020rocket}. Consequently, CNNs are expected to show the best performance in the suppression of the background triggers. A major benefit of CNNs is that they enable model-independent analysis of recorded pulses as one does not have to rely on fitting functions to the data. We will utilize the CNNs as binary classifiers that are trained to distinguish between 1064\,nm photon induced \textit{light pulses} and any other background source induced \textit{dark pulses}. These classifiers are then ensembled to quantitatively study the background sources that are particularly difficult to distinguish from the light pulses and to see whether the CNNs can be used for further background suppression.

The manuscript is organized as follows: in Section~\ref{experimental_data_section} we describe our experimental setup and how the experimental data used in this work was obtained (see Ref.~\cite{RubieraGimeno2024optimizing} for more details). Next, in Section~\ref{CNN_architecture} we present a detailed description of the overall architecture of the considered CNN and explain how we use the experimentally measured data to train it and  evaluate its performance. Details of the CNN's hyperparameter optimization are also presented. The performance of the optimized CNN is analyzed and further fine-tuned in Section~\ref{fine-tuning_optimized_CNN_section}. We then proceed in evaluating the true performance of the model in classifying the light and dark pulses in Section~\ref{performance_of_the_CNN_section} and study the observed background pulses in detail in Section~\ref{background_classification_section}. Finally,  we discuss the background source induced confusion in the CNN training process in Section~\ref{BB-photons_and_training_confusion_section} before summarizing the final conclusions in Section~\ref{conclusions_section}.

\section{Experimental data}
\label{experimental_data_section}
\noindent All of the experimental data was measured using a tungsten-based TES chip fabricated at National Institute of Standards and Technology (NIST). The working point of the voltage-biased TES was set to 30\% of its normal state resistance and its current was monitored by an inductively coupled flux-locked SQUID (manufactured by Physikalisch Technische Bundesanstalt; PTB) with 5\,GHz gain bandwidth product at a sampling rate of 50\,MHz. The photons absorbed by the TES were then detected as pulses in the SQUID output voltage $V_\mathrm{out}(t)$. The TES+SQUID module was operated within a Bluefors SD dilution refrigerator at 25\,mK base temperature. The effective circuit diagram of the used TES setup is illustrated in Fig.~\ref{TES_schematics_Figure_ab}(a), while the layer structure of the TES, optimized for the detection of 1064\,nm photons \cite{Shah2022characterising}, is presented in Fig.~\ref{TES_schematics_Figure_ab}(b). A picture of the TES+SQUID module installed into the dilution refrigerator is shown in Fig.~\ref{TES_schematics_Figure_ab}(c).

In order to recognize the shapes of the 1064\,nm photon $V_\mathrm{out}(t)$ pulses, we first gathered data by illuminating the TES with a highly attenuated 1064\,nm laser source for a total of 5\,s. The laser light was coupled to the TES via a HI1060 single mode optical fiber. During this time interval, a total of 4722 pulses above the 10\,mV trigger threshold were recorded, where each time trace corresponds to a 200\,$\mu$s time window with $10^4$ samples (50\,MHz sampling frequency). The recorded time traces were pre-filtered by discriminating double pulses. This was done by detecting local maxima from the derivative of a time trace based on specific height, prominence and spacing criteria. This left us with 3928 single-photon triggered pulses. We have performed fits in the time domain using a phenomenological function \cite{Meyer2024first, RubieraGimeno2024optimizing}
\begin{equation}
\label{TES_pulse_Ph_Eq}
    V_\mathrm{ph}(t) = -\frac{2A_\mathrm{ph}}{ \mathrm{e}^{\frac{(t_0-t)}{\tau_\mathrm{rise}}} + \mathrm{e}^{-\frac{(t_0-t)}{\tau_\mathrm{decay}}} } + V_0,
\end{equation}
describing a photonic event triggered TES pulse at around $t=t_0-\tau_\mathrm{rise}$. The parameter $A_\mathrm{ph}$ is directly proportional to the amplitude of the pulse, while the pulse shape is determined by the rise and decay times $\tau_\mathrm{rise}$ and $\tau_\mathrm{decay}$, respectively. The obtained distribution of $\tau_\mathrm{rise}$ and $\tau_\mathrm{decay}$ will be used in determining the cuts. The $\chi_\mathrm{ph}^2$-error associated with the performed fit is also considered. In addition, we have also considered a fitting function from \textit{Small Signal Theory} \cite{Irwin2005transition}
\begin{equation}
    \label{TES_pulse_SST_Eq}
    V_\mathrm{SST}(t) =
\begin{cases} 
A_\mathrm{FFT} \cdot \left( \mathrm{e}^{-(t-t_0)/\tau_+} - \mathrm{e}^{-(t-t_0)/\tau_-}\right), & \text{if } t \geq t_0. \\
0, & \text{else},
\end{cases}
\end{equation}
where the parameters $A_\mathrm{FFT}$, $\tau_+$ and $\tau_-$ are analogous to the $A_\mathrm{ph}$, $\tau_\mathrm{rise}$ and $\tau_\mathrm{decay}$ of the phenomenological model introduced above. While fitting Eq.~(\ref{TES_pulse_SST_Eq}) in time domain is unstable, we have performed the fit in frequency domain, in particular due to the simple Fourier transformation of Eq.~(\ref{TES_pulse_SST_Eq}). The obtained fitting parameters were then used to calculate the associated peak height ($\mathrm{V}_\mathrm{min,\,FFT}$) which will also be considered for the filtering of the pulses. This specific parameter was chosen because its associated distribution for light pulses has previously resulted in the highest achieved energy resolution for our TES \cite{Gimeno2024tes, RubieraGimeno2024optimizing}. The $\chi_\mathrm{FFT}^2$-error associated with the fits in frequency domain is also considered for filtering.  

In order to mitigate the effects of scattering processes and nonlinear optical effects (that can alter the wavelength of the photons emitted from the laser) on the training process, we have subjected the 3928 single-photon triggers for minor filtering. This was done by only including pulses whose $\tau_\mathrm{rise}$, $\tau_\mathrm{decay}$ and $\mathrm{V}_\mathrm{min,\,FFT}$ were simultaneously within 0.1\%--99.9\% quantiles of their associated distributions while $\chi_\mathrm{ph}^2$ and $\chi^2_\mathrm{FFT}$ being below 99.9\% quantiles. This resulted in the discrimination of 0.76\% (30) of the filtered triggers, leaving us a total of 3898 pulses that are used for training and evaluating the CNNs.

The filtered 3898 pulses were further post-processed by removing a possible voltage offset by shifting the recorded $V_\mathrm{out}(t)$ values by the average voltage value measured between $t=0\texttt{-}24\,\mu\mathrm{s}$. The waveforms were truncated to a time window of $24\,\mu\mathrm{s}$ (corresponding to 1200~samples) by locating the pulse minimum and including the 300 antecedent and 900 subsequent samples. This ensures that we fully capture the 1064\,nm photon triggered pulses given their average rise and decay times (see Fig.~\ref{pulses_and_PCA_Figure_ab}(a)). Evidently, we strived for minimizing the time window (and the number of samples) in order to limit the computational load for the CNNs. The waveforms were not resampled prior to training of the CNNs in order to prevent any loss of information. For the rest of the manuscript, we will keep referring to these 1064\,nm photon triggered time traces as \textit{light pulses}. The average measured light pulse is presented in Fig.~\ref{pulses_and_PCA_Figure_ab}(a) together with an example of a single pulse in the inset of the figure, further illustrating the baseline noise present in the measurements.  

Right after measuring the light pulses, we proceeded to measure the extrinsic background over a period of two days using the same system configuration except for disconnecting the optical fiber from the laser source and sealing its open end with a metallic cover. A total of 8872 background events exceeding the 10\,mV trigger threshold were observed. While all of these pulses were included as training and evaluation data for the CNNs without making any cuts, they were post-processed in the same way as the light pulses by removing the voltage offsets and clipping the time windows as described above. It should be noted that for the pulses with large rise and decay times, typically originating from intrinsic background sources, the clipped time window can be too narrow to fully represent the entire pulse. Regardless, these pulses would have been in any case very easily distinguishable from the light pulses. We refer to all of the recorded background pulses as \textit{dark pulses} for the rest of the manuscript. The average measured dark pulse is presented in Fig.~\ref{pulses_and_PCA_Figure_ab}(a).
\begin{figure*}[t!]
\begin{center}
\includegraphics[width=13.0cm]{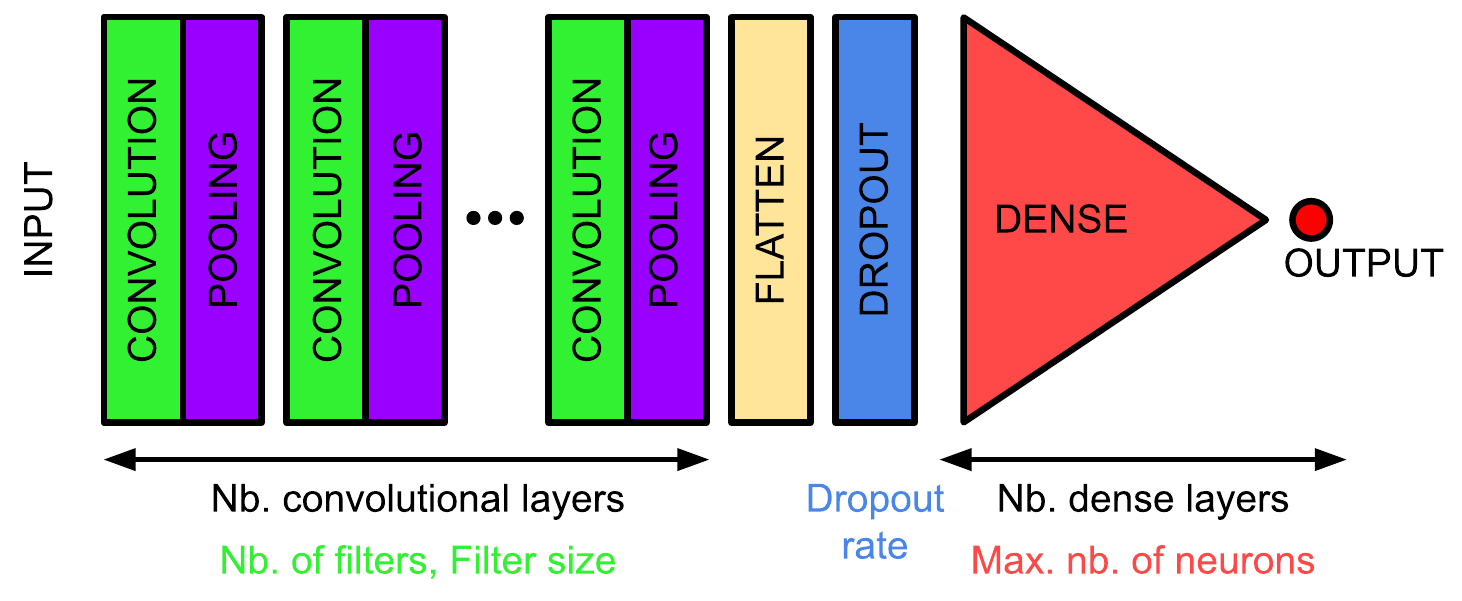}
\caption{A schematic illustration of the basic architecture of the considered CNN and its hyperparameters whose optimization is explicitly addressed (see Table \ref{HP-opt_summary_table}).  }
\label{CNN_schematics_Fig}
\end{center}
\end{figure*}

In summary, after data cleaning we are left with 3898 light pulses and 8872 dark pulses, making the overall size of the dataset 12,770 to be used for the training and evaluation the CNNs. Before proceeding, we want to further characterize the differences between light and dark pulses via Principal Components Analysis (PCA) in order to detect any possible overlap between the resulting light and dark clusters indicating the presence of photonic background. We have done this by associating each pulse with a feature vector assembled from the above introduced fitting parameters as ($\tau_\mathrm{rise}$, $\tau_\mathrm{decay}$, $\chi_\mathrm{ph}^2$, $\mathrm{V}_\mathrm{min,\,FFT}$, $\chi^2_\mathrm{FFT}$), where both $\tau_\mathrm{rise}$ and $\tau_\mathrm{decay}$ are in the units of \si{\micro\second} and $\mathrm{V}_\mathrm{min,\,FFT}$ is in the units of mV. The PCA scatter plot visualizing the projection of these feature vectors for both light and dark pulses onto the two main principal components is presented in Fig.~\ref{pulses_and_PCA_Figure_ab}(b). The obtained loading vector for the first principal component is $\mathbf{w}_\mathrm{PC1} = (9.5\cdot10^{-7},\,9.1\cdot10^{-5},\, 0.99,\, 3.4\cdot 10^{-5},\,7.1\cdot10^{-3} )$, while for the second one $\mathbf{w}_\mathrm{PC2} = ( 1.7\cdot10^{-4},\, 2.8\cdot10^{-2},\, -7.1\cdot10^{-3},\, 1.9\cdot10^{-3},\, 0.99)$. The primary modes of variance are thus associated with the $\chi_\mathrm{ph}^2$ and $\chi_\mathrm{FFT}^2$ errors, reflecting the fact that the used fitting functions describe photonic triggers while majority of the dark counts originate from intrinsic non-photonic background, that is easily distinguishable. As expected, the light pulses are tightly clustered in one spot while the dark pulses are much more spread out. Regardless, one can observe significant overlap between the light and dark pulses as illustrated in the inset of Fig.~\ref{pulses_and_PCA_Figure_ab}(b), most likely originating from fiber coupled black-body radiation \cite{Gimeno2025simulation}. We will analyze this later in the paper aided by the CNN ensembles. 

\section{CNN Architecture}
\label{CNN_architecture}
\noindent The basic architecture of the CNN considered in this manuscript resembles the generally used structure for image classifying tasks \cite{Krizhevsky2012imagenet, Lecun2015deep}. As illustrated in Fig.~\ref{CNN_schematics_Fig}, the CNN consists of pairs of i) convolutional and average pooling layers followed by ii) flatten and dropout layers connected to iii) dense layers with gradually decreasing sizes ultimately ending up to a single output neuron. As typical for binary classifiers, we use \textit{binary cross-entropy (log-loss)} as the loss function to guide the training process. The weights of the model are initialized using the \textit{Glorot uniform} approach \cite{Glorot2010understanding} and updated during training following the \textit{Adaptive Moment Estimation (Adam)} algorithm \cite{Kingma2014adam}. 

In order to limit the size of the search space for hyperparameter optimization, we fix the activation functions associated with the convolutional layers to \textit{tanh} while using \textit{ReLu} for the dense layers. This combination was observed to clearly result in best performance in our initial tests of the model. We further require that all of the convolutional layers share the same \textit{number of filters}, \textit{filter size}, and \textit{pool size}. We fix the pool size to 2, limiting the maximum number of convolutional layers in the considered architecture to 10. The structure of the dense layers is limited by requiring that the number of neurons must always drop by half in the following hidden layer. This leaves only the \textit{maximum number of neurons within the first layer} and the \textit{number of hidden layers} as the only hyperparameters to be optimized for the dense part of the CNN. On top of the architectural hyperparameters, we address the optimization of the \textit{dropout rate}, \textit{number of epochs} and \textit{batch size}. A summary of the considered search space for the hyperparameter optimization is presented in Table \ref{HP-opt_summary_table}. The CNN is implemented using a high-level neural network API Keras version 2.12.0 \cite{Chollet2015keras} with a TensorFlow version 2.12.1 backend \cite{Tensorflow2015-whitepaper}. 

\begin{figure}[t!]
\begin{center}
\includegraphics[width=8.0cm]{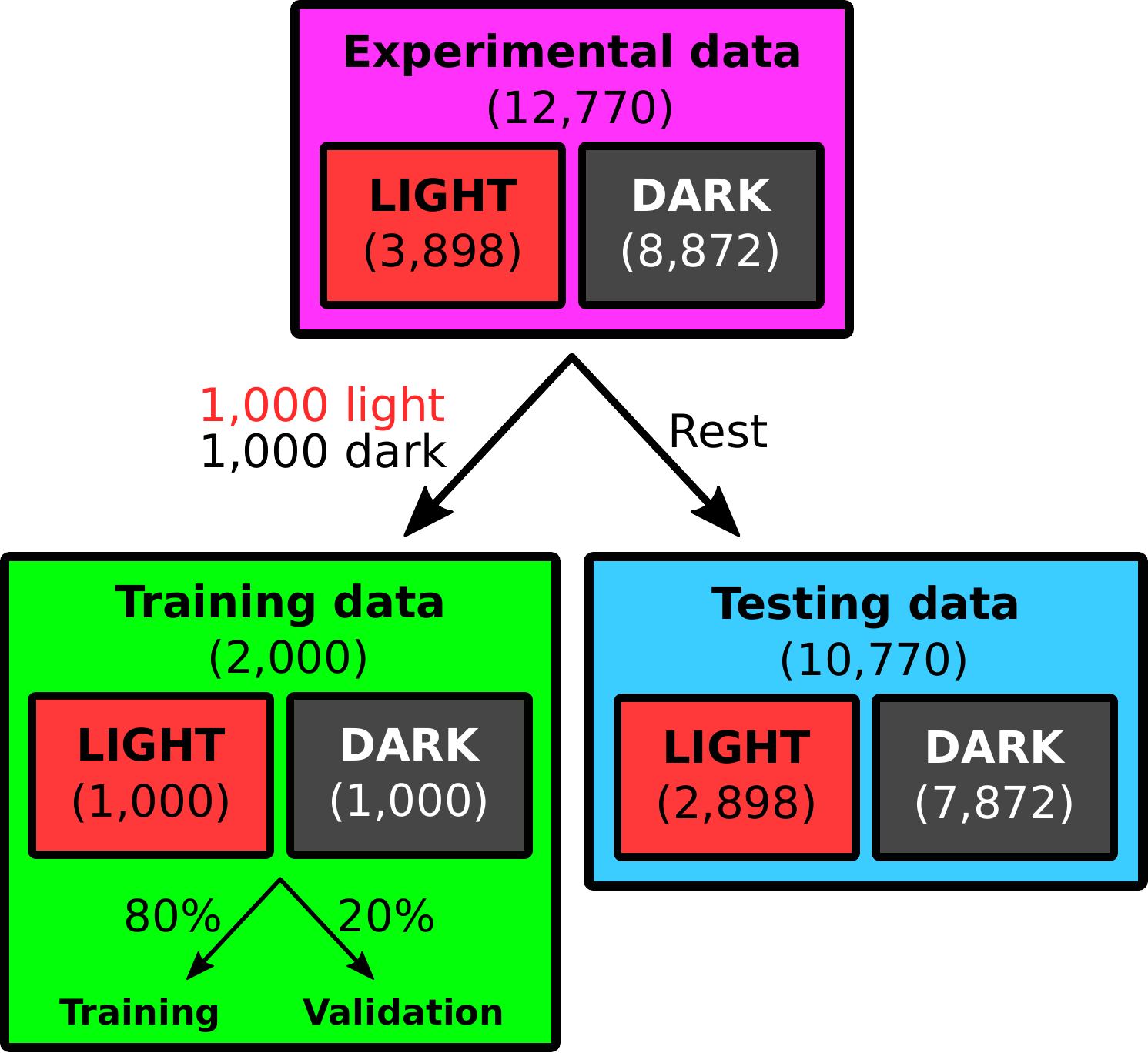}
\caption{A schematic illustration of the division of the dataset into training and testing data. The training set was further divided 80\%-20\% into training and validation sets, where the validation set was to evaluate the performance of the CNN during the training process. }
\label{training_and_testing_data_schematics_Fig}
\end{center}
\end{figure}

\subsection{Training process}
\label{training_process_section}
\noindent The model is trained using 1000 light and dark pulses, respectively, resulting in an overall training set size of 2000 pulses. It should be noted that the training set is perfectly balanced between light and dark pulses, making it optimal for training. The training set is further split 80\%--20\% into training and validation sets, where the training set is used for the actual training of the model while the evaluation set is used to guide the training process via minimization of the \textit{binary cross-entropy} used as the loss function. The division of the dataset into training and testing set is schematically illustrated in Fig.~\ref{training_and_testing_data_schematics_Fig}.

\subsection{Performance evaluation}
\label{performance_evaluation_section}
\noindent The performance of the model is evaluated against the rest of the dataset that was not dedicated for training as illustrated in Fig.~\ref{training_and_testing_data_schematics_Fig}, consisting of 2898 light pulses and 7872 dark pulses. With our main interest towards the use of the CNN in the ALPS~II experiment, we follow the approach of our previous work (Ref.~\cite{Meyer2024first}) and evaluate the performance of the model during hyperparameter optimization based on the detection significance given by \cite{Bityukov1998new, Bityukov2000observability}
\begin{equation}
\label{S_Eq}
    S = 2 \sqrt{T_\mathrm{obs}} \cdot \left( \sqrt{\epsilon_\mathrm{d} \epsilon_\mathrm{a} n_\mathrm{s} + n_\mathrm{b}} - \sqrt{n_\mathrm{b} } \right),
\end{equation}
where $T_\mathrm{obs}=518\,\mathrm{h}$ is the observation time of the experiment (as used in Ref.~\cite{Meyer2024first}), $n_\mathrm{s}=2.8\cdot 10^{-5}\,\mathrm{Hz}$ is the assumed signal (1064\,nm photon) rate and $\epsilon_d=0.5$ is the pessimistically evaluated detection efficiency taking into account all losses associated with the experimental setup \cite{Meyer2024first}. The only analysis method dependent parameters are the closely related background rate ($n_\mathrm{b}$) and analysis efficiency ($\epsilon_\mathrm{a}$). The $\epsilon_\mathrm{a}$ is simply calculated as the percentage of correctly classified light pulses (true positive). The $n_\mathrm{b}$ on the other hand is calculated from the number of misclassified dark pulses ($N_\mathrm{mdp}$, false positives). Since the total of 8872 extrinsic background pulses were measured over a time period of 2\,d, the used testing set containing the subset of 7872 dark pulses effectively corresponds to $(7872/8872)\cdot 2\,\mathrm{d} \approx 1.77\,\mathrm{d}$ time period. The effective background rate can thus be estimated from the number of misclassified dark pulses (false positives) as $N_\mathrm{mdp}/1.77\,\mathrm{d}$. It should be pointed out that the $S$ score is a threshold ($\mathrm{Th.} \in [0,1]$ corresponding to dark--light, respectively) dependent metric. Consequently, all the reported values of $S$ in this manuscript have been obtained after optimizing the threshold to maximize its value.

While the $S$ score will be used to determine the optimal combination of hyperparameters in the following section, we will later also evaluate the trained CNNs using the $F_1$ score ($F_1=2\times \mathrm{Precision}\times \mathrm{Recall} /( \mathrm{Precision}+\mathrm{Recall} )$ which balances between precision (TP/(TP+FP)) and recall (TP/(TP+FN)) thus making it a commonly used evaluation metric for biased datasets (TP=True~Pos., FP=False~Pos., FN=False~Neg.). The $F_1$ score describes the true classification performance of the CNN better than the $S$ score by directly measuring how well the CNN is able to correctly classify the light pulses while avoiding the misclassification of dark pulses. Thus, we will utilize the $F_1$ score in particular in Section~\ref{background_classification_section}, where we study the nature and origins of the background pulses which the CNNs are struggling to classify correctly in more detail. The $F_1$ score is also a threshold dependent metric and its reported values in later sections have been obtained after the threshold optimization. It should be pointed out that the threshold optimization has to be done separately for the $S$ and $F_1$ scores. 
\begin{figure}[t!]
\begin{center}
\includegraphics[width=6.0cm]{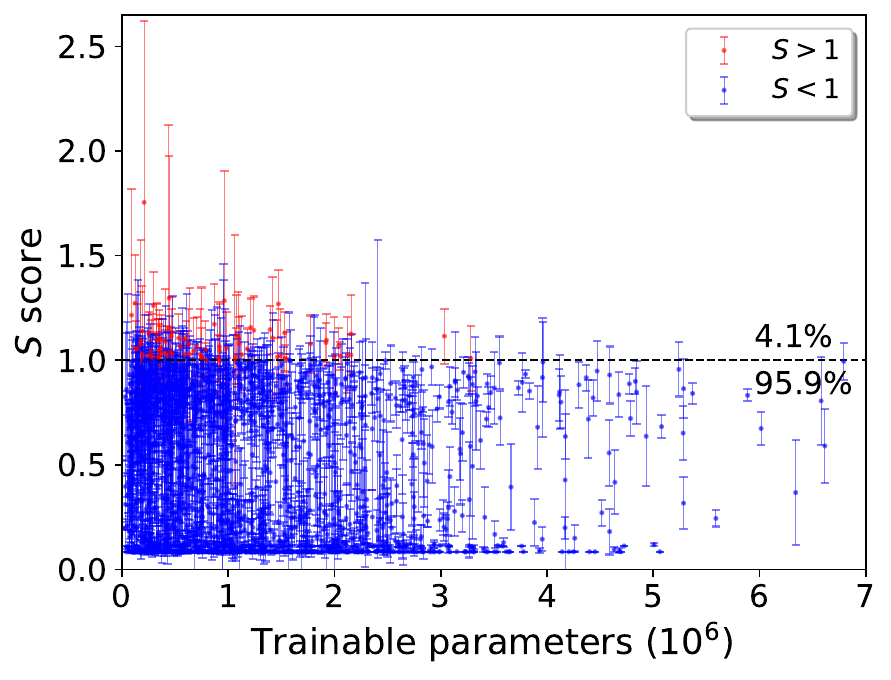}
\caption{The evaluated average $S$ scores as a function of number of trainable parameters for the associated CNN. The error bars correspond to standard deviations associated with 5 evaluations of the CNN with different training and testing sets. The dashed vertical line points to the limit $S=1$ above which the points are colored by red (4.1\%) and below as blue (95.9\%).}
\label{hpopt_Fig}
\end{center}
\end{figure}

\begin{table}[b!]
\centering
\begin{adjustbox}{max width=\textwidth}
\begin{tabular}{ c|c|c|c|c } 
\textbf{Hyperparameter} & \textbf{Optimization range 1} & \textbf{Optimum 1} & \textbf{Optimization range 2} & \textbf{Optimum 2} \\
\hline
Nb. of conv. layers & 3--10 & 6 & 4--7 & 7 \\
Nb. of filters & 20--150 & 45 & 20--60 & 40   \\
Kernel size & 3--20 &  12 & 5--15 & 7 \\
Dropout rate & 0--0.2$^\dagger$ & 0.18 & 0.05--0.2 (step: 0.01)& 0.07  \\
Nb. of dense layers & 1--10  & 3 & 3 (fixed) & 3 \\
Max nb. of neurons & 100--300 & 188 & 188 (fixed) & 188 \\
Learning rate & $10^{-5}$--$10^{-3}$$^\dagger$ & $5.2\cdot10^{-4}$ & $5.2\cdot10^{-4}$ (fixed) & $5.2\cdot10^{-4}$ \\
Epochs & 5--20 &  10$^*$ & 20 (fixed) & 20 \\
Batch size & 32--128 & 99 & 99 (fixed) & 99 \\
\hline
\multicolumn{1}{c}{}& &  $\langle S \rangle = 1.26\pm 0.16$ & & $\langle S \rangle = 1.24\pm 0.05$ \\
\end{tabular}
\end{adjustbox}
\caption{The initial search space for the hyperparameter optimization (\textit{Optimization range~1}) of the CNN using 2000 iterations of random search. The activation functions of the convolutional and dense layers were fixed to \textit{tanh} and \textit{ReLu}, respectively. The presented optimum (\textit{Optimum 1}) corresponds to the maximum obtained average detection significance $\langle S \rangle$ (see details in Section~\ref{performance_evaluation_section}) for an ensemble of 5 CNNs trained and evaluated with differently (randomly) divided training, validation and testing sets, while the weight initialization for the CNN was fixed. The values of $S$ were calculated using the optimal threshold that maximizes its value. After finding the initial optimum, the search space was narrowed down (\textit{Optimization range~2}) and the alternative optimum (\textit{Optimum~2}) was found using 5000 iterations of random search, covering 20.8\% of the total search space. Narrowing down the search space did not result in improved $\langle S \rangle$ and consequently the combination of hyperparameters used for the rest of the study was fixed to the initial \textit{Optimum~1}. $^*$The number of epochs was later increased to 20 (see Section~\ref{fine-tuning_optimized_CNN_section}). $^\dagger$Continuous range}
\label{HP-opt_summary_table}
\end{table}

\begin{figure*}[t!]
\begin{center}
\includegraphics[width=13.0cm]{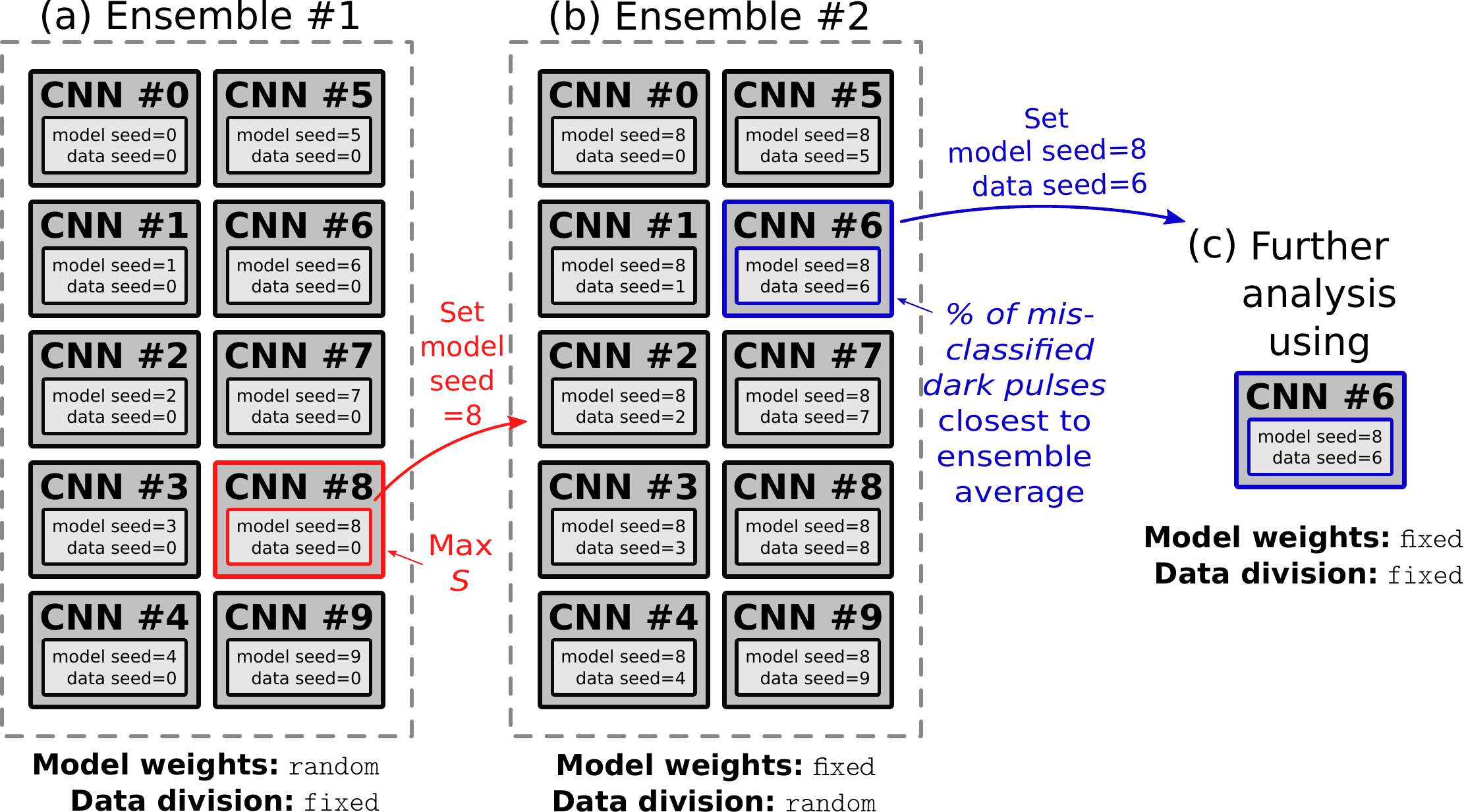}
\caption{A schematic illustration of the CNN ensembles (unweighted averaging ensembles) used in this study. (a) The ensemble used in Section~\ref{fine-tuning_optimized_CNN_section} where the CNNs are trained and evaluated with exactly the same datasets but the initialization of their weights differ between each other. (b) The ensemble used in Section~\ref{performance_of_the_CNN_section} and \ref{background_classification_section}. The model weights were fixed by seeding them equivalently to that of the CNN in the \textit{ensemble \#1} that achieved the highest $S$ score (Eq. (\ref{S_Eq})). Only the dataset division into training and testing sets was randomized for this ensemble. (c) A single CNN trained with data divided equivalently to that of the CNN in the \textit{ensemble \#2} whose percentage of misclassified dark pulses was closest to the average of the ensemble. This model was used in the latter part of Section~\ref{background_classification_section}.}
\label{CNN_ensemble_figure_abc}
\end{center}
\end{figure*}

\subsection{Hyperparameter Optimization}
\label{hyperparameter_optimization}
\noindent The hyperparameters of the CNN architecture introduced in Section~\ref{CNN_architecture} are optimized by 2000 iterations of \textit{random search}, i.e. by training a total of 2000 models using randomized combinations of hyperparameters and choosing the one with highest evaluation metrics. The search space for the considered hyperparameters is presented in Table~\ref{HP-opt_summary_table}; \textit{Optimization~range~1}. In order to reduce the susceptibility of the evaluated performance towards the random division of the dataset into training and testing sets (Fig.~\ref{training_and_testing_data_schematics_Fig}), each iterated combination of hyperparameters is evaluated using 5 CNNs trained and tested with different, randomly divided, datasets as described in Sections \ref{training_process_section} and \ref{performance_evaluation_section}. The initial weights of the CNNs were fixed between the iterations of the random search. 

The evaluated average $S$ scores ($\langle S \rangle$) and their associated standard deviations ($\sigma_S$) as a function of trainable parameters in the CNN are illustrated in Fig.~\ref{hpopt_Fig}. The highest $S$ scores are clearly associated with CNNs with smaller number of trainable parameters, making the training process more efficient. We determine the optimal combination of hyperparameters for the CNN that maximizes $\langle S \rangle - \sigma_S$, under the constraint of limiting the maximum number of trainable parameters to $0.5 \cdot 10^6$. The chosen optimal model has a total of 297,057 parameters and reached $\langle S \rangle = 1.26\pm 0.16$. The associated hyperparameters are presented in Table~\ref{HP-opt_summary_table};~\textit{Optimum~1}. 

While the search space associated with the above described hyperparamter optimization is huge when compared with the 2000 iterations of random search, one might argue that the performance of the CNN would be still limited by the non-optimal combination of hyperparameters. To address this, we have further performed fine-optimization by significantly narrowing down the search space. Considering the convolutional layers the most critical ones when it comes to the performance of the model, we fixed all the parameters associated with the dense layers to the previously found optima, together with learning rate and batch size. We fixed the number of epochs to 20, as we later found that it had room for improvement from the previously optimized value of 10 (see Section~\ref{fine-tuning_optimized_CNN_section}). Meanwhile, we narrowed down the ranges of parameters associated with the convolutional layers so that they still include the previously found optima (see Table~\ref{HP-opt_summary_table};~\textit{Optimization~range~2}). The size of the new search space is 24,000 and we optimized the associated hyperparameters via 5000 iterations of random search, covering a total of 20.8\% of the whole search space. This coverage is generally considered adequate for finding a near-optimum combination of hyperparameter \cite{Bergstra2012random}. Regardless, the found optimum is associated with $\langle S \rangle = 1.24\pm 0.05$ aligning with the initial random search within the uncertainty. Thus, we conclude that the initially found optimal combination of hyperparameters must be close to optimum and the choice of hyperparameters is unlikely to significantly limit the performance of the CNN. Thus, we will proceed with the initially optimized model (Table~\ref{HP-opt_summary_table}; \textit{Optimum~1}).

\section{Fine-tuning optimized CNN}
\label{fine-tuning_optimized_CNN_section}
\begin{figure*}[t!]
\begin{center}
\includegraphics[width=6.0cm]{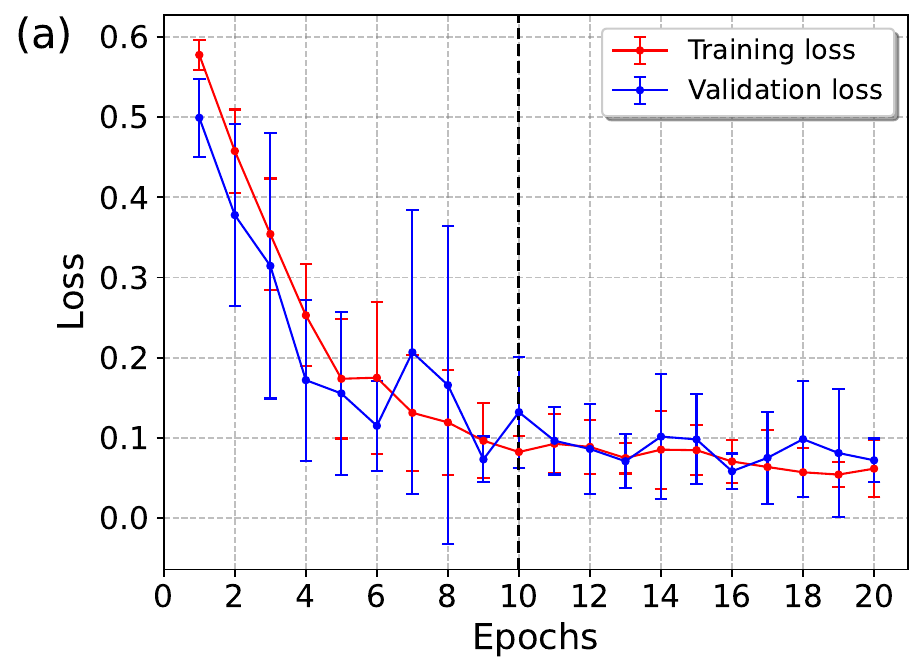}
\includegraphics[width=6.0cm]{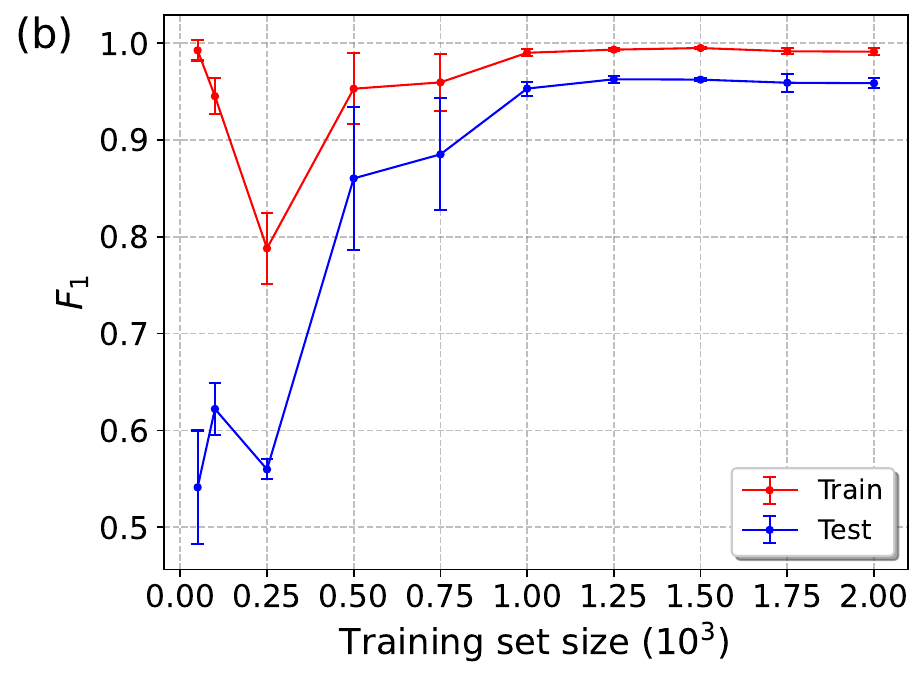}
\caption{Different learning curves associated with the CNN \textit{ensemble \#1}; (a) The average loss (binary cross-entropy) as a function of epochs for the whole ensemble of 10 CNNs. The dashed vertical line indicates the epochs=10 obtained from the initial hyperparameter optimization (see Table~\ref{HP-opt_summary_table}). This was then increased to 20 in order to identify underfitting or overfitting from the associated figure. (b) The $F_1$ score as a function of training set size for the CNN associated with highest $F_1$ (also highest $S$) within the ensemble. The error bars correspond to the standard deviation associated with 3 statistical repetitions in the CNN training, between which the used training data was shuffled. The used testing data was kept constant throughout the process.} 
\label{CNN_ensemble1_learning_curves_ab}
\end{center}
\end{figure*}

\noindent Next, we will proceed investigating and fine-tuning the previously optimized CNN (Table~\ref{HP-opt_summary_table}; \textit{Optimum 1}). In order to identify possible underfitting or overfitting, we increase the number of epochs from the optimized value of 10 up to 20. We train an ensemble of 10~CNNs (unweighted averaging ensemble) with fixed datasets but randomly initialized weights. This CNN~ensemble will be referred to as \textit{ensemble~\#1} (see Fig.~\ref{CNN_ensemble_figure_abc}(a)). We noticed that the performance of CNNs is susceptible towards the random initialization of its weights. This is manifested as a high standard deviation in the $\langle S \rangle = 0.98\pm 0.22$ (average threshold: $\langle \mathrm{Th.} \rangle = 0.98 \pm 0.012$) for \textit{ensemble~\#1} (see Fig.~\ref{CNN_ensemble_figure_abc}(a)). While the weight initialization is internally stochastic, we have initialized the \textit{ensemble~\#1} weights in a deterministic manner via specified random seeds. This enables us to optimize the weight initialization of the CNN associated with the chosen seed for reproducible results. The observed sensitivity towards weight initialization is particularly common for CNNs associated with dropout layers. Similar observations have been made in other CNN based machine learning implementations \cite{Ismail2020inceptiontime}. The high values of optimal thresholds with respect to the $S$ score reflect the importance of decreasing the background rate at the cost of analysis efficiency (see Eq.~(\ref{S_Eq})).

We also evaluate \textit{ensemble~\#1} using the $F_1$ score (see \ref{performance_evaluation_section}) reflecting the sole classification performance of the ensemble. We obtain $\langle F_1 \rangle = 0.96 \pm 0.006$ ($\langle \mathrm{Th.} \rangle = 0.63\pm 0.34$), indicating reduced conservativity in the classification of light pulses when compared with the $S$ score. This demonstrates the importance of threshold optimization for different metrics. While the high threshold values resulting in low background rates at the expense of analysis efficiency is preferable for the $S$ score, the best performance for the CNN, in terms of correctly classifying the light and the dark pulses, is obtained at much lower thresholds where the $F_1$ score is maximized.

\begin{table*}[b!]
\small
    \centering
    \begin{tabular}{c|c|c}

         & \textbf{Avg. Detection Significance} & \textbf{Avg. \textit{F}}$_\mathbf{1}$ \textbf{Score}     \\ 
         & $\langle S \rangle = 0.95 \pm 0.23$  &  $\langle F_1 \rangle = 0.961 \pm 0.005 $ \\ 
        \hline
        \hline
        Avg. Optimal Threshold &   $0.98 \pm 0.01$ & $ 0.68\pm 0.31 $   \\
        ($\langle \mathrm{Th.} \rangle$)  &  & \\ 
        \hline
        Avg. Analysis Efficiency &   $75.3\% \pm 14.3\%$  &  $97.5\%  \pm 0.4\%$    \\
        ($\langle \mathrm{True\,\,Positives} \rangle = \langle \epsilon_\mathrm{a} \rangle$)   & &  \\ 
        \hline
        Avg. misclassified dark pulses &  $0.6 \%\pm 0.6 \%$  &  $2.00\% \pm 0.4 \%$    \\
        ($\langle \mathrm{False\,\,Positives} \rangle$)  &  &  \\ 
        \hline
        Background rate ($n_\mathrm{b}$)& $0.33\,\mathrm{mHz}\pm 0.33\,\mathrm{mHz}$ & $1.0\,\mathrm{mHz}\pm0.2\,\mathrm{mHz}$\\

    \end{tabular}
    \caption{A summary of the evaluated performance of the CNN ensemble~\#2 (Fig.~\ref{CNN_ensemble_figure_abc}(b)) in terms of detection significance and $F_1$ score (Section~\ref{performance_evaluation_section}). The detection significance has been calculated using Eq.~(\ref{S_Eq}) with fixed observation time $T=518\,\mathrm{h}$, detection efficiency $\epsilon_\mathrm{d}=0.5$ and signal rate $\epsilon_\mathrm{s}=2.8\cdot10^{-5}\,\mathrm{Hz}$ based on the current realistic limitations for the ALPS~II experiment \cite{Meyer2024first}. The reported background rate ($n_\mathrm{b}$) is calculated from the average percentage of misclassified dark pulses (false positives) by dividing its value by the effective observation time for the extrinsic background as explained in Section~\ref{performance_evaluation_section}. The reported confidence intervals correspond to standard deviations associated with the ensemble average. The reported standard deviation for the avg. misclassified dark pulses is slightly lower than the average value, but is rounded up.}
    \label{ensemble2_evaluation_table}
\end{table*}

The \textit{ensemble~\#1} average learning curves with respect to number of epochs are presented in Fig.~\ref{CNN_ensemble1_learning_curves_ab}(a). Note, that the number of epochs was increased from 10 (obtained from initial hyperparameter optimization) to 20 in order to identify possible under or overfitting. The learning curve in Fig.~\ref{CNN_ensemble1_learning_curves_ab}(a) indeed indicates slight underfitting and increasing the number of epochs to 20 flattens the learning curves without resulting in overfitting. Thus, the number of epochs will be fixed to 20 for the rest of the manuscript. The learning curves with respect to training set size, calculated for the best performing model within the \textit{ensemble \#1} in terms of the $F_1$ score, are presented in Fig.~\ref{CNN_ensemble1_learning_curves_ab}(b). No underfitting or overfitting is observed from these curves, indicating that the used training set of size 2000 (1000~light, 1000~dark) is well suited for training the models. 

The best performing CNN within the \textit{ensemble \#1} achieved the highest $S=1.4$ using a threshold of $\mathrm{Th.} = 0.97$. We will proceed with initializing the weights of the CNNs used in the following sections according to this specific model. 

\section{Performance of the CNN}
\label{performance_of_the_CNN_section}
\noindent After the optimization of the CNN's hyperparameters (Section~\ref{hyperparameter_optimization}) together with additional optimization of number of epochs and weight initialization (Section~\ref{fine-tuning_optimized_CNN_section}), we now proceed in studying how well the CNN actually performs for its designated purpose as a binary classifier for the light and dark pulses. Here, it is important to note that the $S$- and $F_1$ scores evaluated for the trained CNN can be expected to be susceptible to the random division of the dataset into training, validation and testing sets. This is particularly true in the case where a small subset of mislabeled pulses (e.g. target value for a dark pulse\,=\,1) exists. Consequently, we again train an ensemble of 10~CNNs (unweighted averaging ensemble) and study its average performance on classifying the pulses. We will keep referring to this as \textit{ensemble~\#2} that is schematically illustrated in Fig.~\ref{CNN_ensemble_figure_abc}(b). Unlike for the \textit{ensemble~\#1} studied in the previous section, the weights of all the models in the \textit{ensemble~\#2} are initialized similarly based on the results of the previous Section~\ref{fine-tuning_optimized_CNN_section}. However, all the CNNs in \textit{ensemble~\#2} are trained and evaluated with different randomly divided datasets according to Fig.~\ref{training_and_testing_data_schematics_Fig}. 

The average evaluation metrics for the CNN \textit{ensemble~\#2} is listed in Table~\ref{ensemble2_evaluation_table}. As expected, the CNNs turned out to be susceptible to the random division of the datasets with ensemble average evaluation metrics $\langle S \rangle = 0.95 \pm 0.23$ ($\langle \mathrm{Th.} \rangle = 0.98\pm 0.01 $) and $\langle F_1 \rangle = 0.961 \pm 0.005 $ ($\langle \mathrm{Th.} \rangle = 0.68\pm 0.31 $) having rather large standard deviations. The highest $S$ scores are obtained at much higher values of thresholds when compared with the $F_1$ scores. As seen in Table \ref{ensemble2_evaluation_table}, the associated differences in analysis efficiency and number of misclassified dark pulses ($\sim$background rate) reflect the importance of background suppression at the expense of analysis efficiency for maximizing $S$. 

\begin{figure}[t!]
\begin{center}
\includegraphics[width=6.0cm]{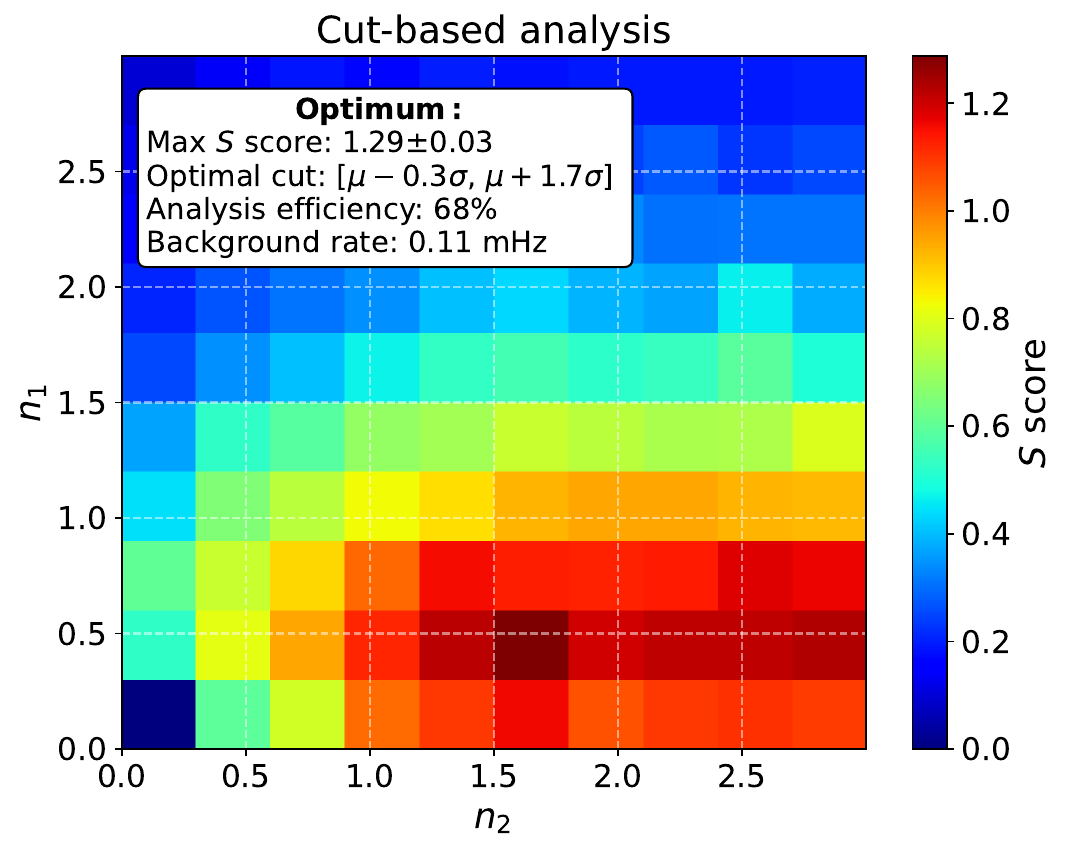}
\caption{The detection significance as a function of the cut range associated with the $\mathrm{Ph}_\mathrm{FFT}$ calculated based on the fitting parameters obtained in frequency domain analysis. The cut range is defined as $[\mu_\mathrm{m} - n_1\cdot \sigma,\,\mu_\mathrm{m} + n_2\cdot \sigma]$, where $n_1,n_2 \in 0, 1/3, \dots, 3$ and the $\mu_\mathrm{m}\approx-16.33\,\mathrm{mV}$ and $\sigma\approx1.25\,\mathrm{mV}$ are obtained by fitting a skewed Gaussian to the distribution of the peak heights for the light pulses. All of the detection significances were determined as the average of 5 $S$ scores calculated for randomly chosen testing sets. The confidence interval of the $S$ score corresponds to standard deviation associated with the average.}
\label{cut-based_analysis_Fig}
\end{center}
\end{figure}

\subsection{Comparison to Cut-Based Analysis}
\label{comparison_to_cut-based_analysis_section}
\noindent Next, we want to compare the above presented pulse analysis using the CNN \textit{ensemble \#2} with the traditional (non-ML) cut-based analysis. In the cut-based analysis the pulse is classified based on its associated fitting parameters $\tau_\mathrm{rise}$, $\tau_\mathrm{decay}$, $\chi_\mathrm{Ph}^2$, $\mathrm{Ph}_\mathrm{FFT}$ and $\chi^2_\mathrm{FFT}$ introduced in Section~\ref{experimental_data_section}. In order for a pulse to be classified as light, all of the pulse parameters must simultaneously lie within $[\mu_\mathrm{m} - 3\cdot \sigma, \mu_\mathrm{m} + 3\cdot \sigma]$ for the parameter specific $\mu_\mathrm{m}$ and $\sigma$. The only exception is with the $\mathrm{Ph}_\mathrm{FFT}$, for which the cut range $[\mu_\mathrm{m} - n_1\cdot \sigma, \mu_\mathrm{m} + n_2\cdot \sigma]$ will be rigorously optimized for $n_1,n_2 \in 0, 1/3, \dots, 3$. These cut ranges in terms of $\mu_\mathrm{m}$ and $\sigma$ are determined based on the fits of skewed Gaussians to the parameter distributions.

However, in order to make the comparison between the performance of the CNN \textit{ensemble \#2} and cut-based analysis fair, we determine the cut ranges for the associated pulse parameters based on 1000 randomly selected light pulses. In analogy to machine learning, determining the cut region corresponds to the training of the model. The rest of the light data together with the whole extrinsic dataset is then used as testing data to evaluate the $S$ score. It should be noted that the triggers used for determining the cuts (training) and evaluating the $S$ score (testing) correspond exactly to the pulses used for training and evaluating the CNNs. Thus, the used light pulses have been filtered as described in Section~\ref{experimental_data_section}. 

The optimization of the cut range for $\mathrm{Ph}_\mathrm{FFT}$ is illustrated in Fig. \ref{cut-based_analysis_Fig}, where every calculated $S$ score represents the average of 5 $S$ scores calculated for randomly selected testing data. The cut-based analysis results in the maximum detection significance of $S=1.29\pm0.03$ achieved with the optimal cut $[ \mu - 0.3\sigma,\, \mu + 1.7\sigma ]$. The obtained average score is approximately 36\% higher when compared with the average $\langle S \rangle = 0.95 \pm 0.23$ achieved by the CNN~\textit{ensemble~\#2}.

While being outperformed by the cut-based analysis, we argue that the CNN's performance is limited by the measured extrinsic dataset (containing presumed dark pulses) that has been distorted by actual light pulses (outliers). That is, the dark data contains a subset of pulses that actually correspond to 1064\,nm photon induced triggers. Such dataset distortion would cause confusion in the training of the CNNs thus limiting their performance. Of course, the presence of the near-1064\,nm black-body photons in the dark dataset also has detrimental effect on the $S$ score calculated using the cut-based analysis, but since the dataset used to calculate the $S$ scores using the CNNs and cut-based analysis are the same, their comparison with each other is fair. In the following sections, we will focus on investigating the background pulses that were previously presumed as dark and based on the results, quantitatively address how the near-1064\,nm photon black-body photon induced distortion in the dark dataset results in training confusion.

\section{Background classification}
\label{background_classification_section}
\noindent We will use the CNN \textit{ensemble~\#2} from Section~\ref{performance_evaluation_section} to study the nature and origin of the remaining background set by the misclassified dark pulses (false positives). To do this, one needs to work with the $F_1$ score that quantifies how well the model classifies light pulses as light while avoiding misclassifying dark pulses and light. With the associated optimal thresholds, the \textit{ensemble~\#2} correctly classifies $97.5\%\pm 0.4\% $ of the light pulses (analysis efficiency) while misclassifying $2.00\% \pm 0.4\%$ of the dark pulses as light (Table \ref{ensemble2_evaluation_table}). The latter corresponds to the average of 157 misclassified dark pulses. It is likely that this subset of dark pulses is triggered by photonic events, making them difficult to distinguish from the light pulses. In fact, we have previously concluded that the limiting background source for our TES is fiber coupled near--1064\,nm black-body photons which are indistinguishable from the light pulses given the energy resolution of the TES \cite{Gimeno2023tes, Gimeno2024tes, Gimeno2025simulation}. In order to confirm this, we begin by comparing the observed effective rate of the assumed near-1064\,nm black-body background to theoretical predictions. Using the 1.77\,d observation time derived in Section~\ref{performance_evaluation_section}, the observed background rate is $n_\mathrm{b}=157/1.77\,\mathrm{d}$ that equals approximately $1.02\,\mathrm{mHz} \pm 0.2\,\mathrm{mHz}$

\begin{figure}[t!]
\begin{center}
\includegraphics[width=6.0cm]{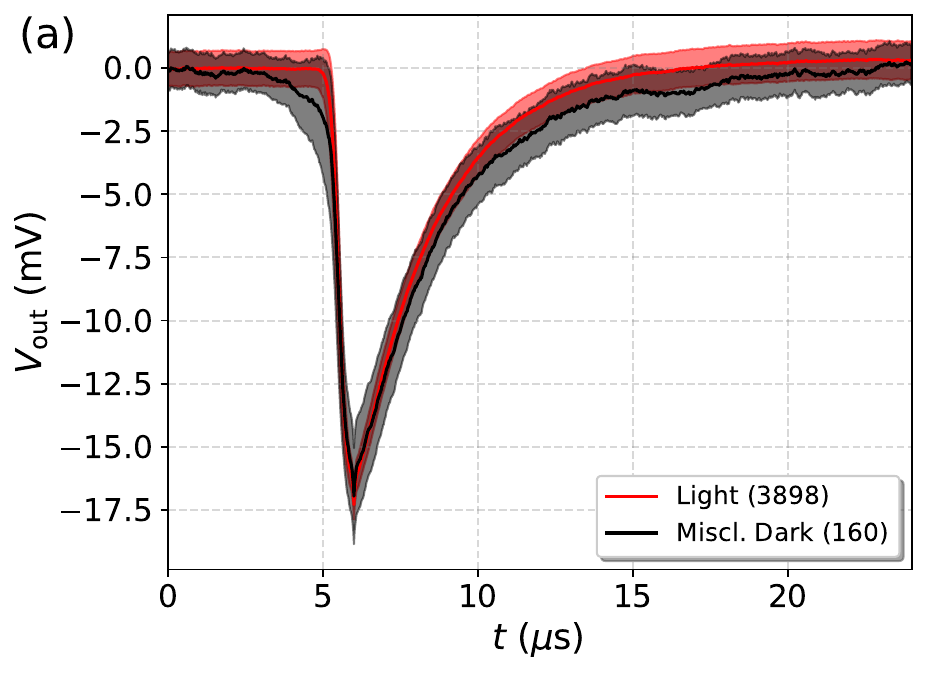}
\includegraphics[width=6.0cm]{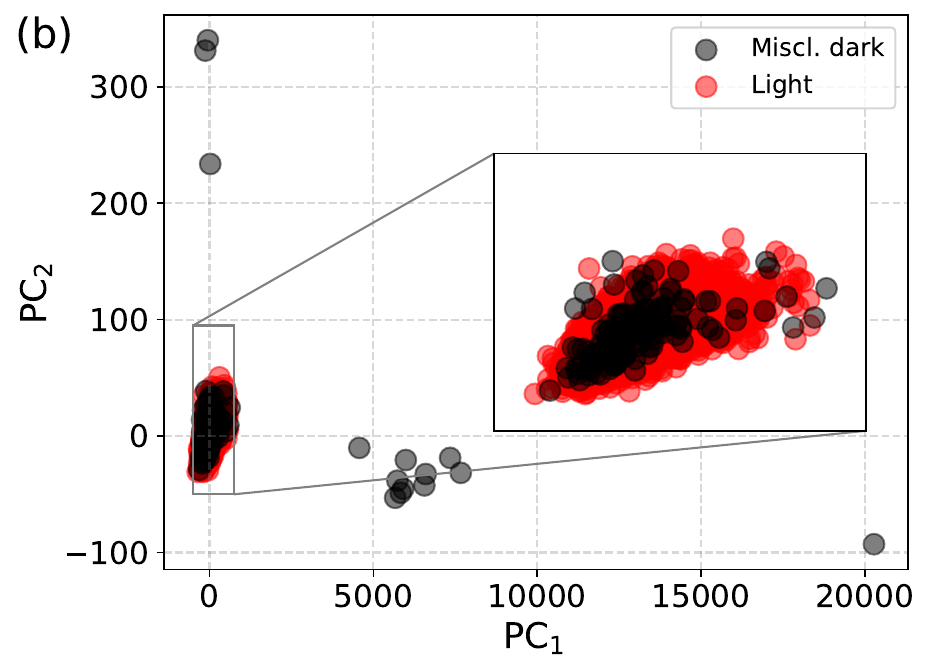}
\caption{(a) The average light pulse used for training and testing the CNN together with the average misclassified dark pulse. (b) PCA scatter plot showing the projection of feature vectors ($\tau_\mathrm{rise}$, $\tau_\mathrm{decay}$, $\chi_\mathrm{Ph}^2$, $V_\mathrm{min,\,FFT}$, $\chi^2_\mathrm{FFT}$) onto the first two principal components (PC$_1$ and PC$_2$) for all of the light (true positives) and misclassified dark pulses (false positives). The loading vectors associated with the principal components are $\mathbf{w}_\mathrm{PC1}=(2.9\cdot 10^{-4},\, 1.0\cdot 10^{-3},\,0.99, -4.6\cdot 10^{-5},\,5.5\cdot 10^{-3})$ and $\mathbf{w}_\mathrm{PC2}=(-1.2\cdot 10^{-3},\, 6.0\cdot 10^{-3},\, 5.5\cdot 10^{-3}, -1.4\cdot 10^{-2},\,0.99)$, again suggesting that the primary modes of variance are associated with the $\chi_\mathrm{ph}^2$ and $\chi_\mathrm{FFT}^2$ errors. }
\label{misclassified_dark_pulses_Fig-ab}
\end{center}
\end{figure}

The expected rate of 1064\,nm black-body photons can be theoretically estimated from the Planck's spectrum
\begin{equation}
\label{Plancks_Spectrum_Eq}
    \Dot{N} = \int d\Omega \int dA \int_{E_1}^{E_2}  \frac{2}{h^3 c^2} \cdot \frac{E^2}{\mathrm{e}^{E/kT}-1}\,dE,
\end{equation}
where the first two integrals represent the solid angle ($\Omega$) and the area ($A$) over which the black-body radiation can enter the optical fiber, and $h$ is Planck's constant, $c$ is the speed of light, $T$ is the temperature of the black-body radiation source and $k$ is Boltzmann's constant. The integrals over $d\Omega$ and $dA$ are purely geometrical and can be estimated using the supplier provided specs of the used HI-1060 single mode fiber; numerical aperture $\mathrm{NA}=0.14$ and core radius $R=3.1\,\mu\mathrm{m}$. The solid angle is calculated as $\Omega = 2\pi\cdot (1 - \mathrm{cos} (\theta))$, where the acceptance angle of the fiber is $\theta = \mathrm{sin}^{-1}(\mathrm{NA})$. This results in $\Omega=0.062$. The corresponding area is simply $A=\pi R^2 = 3\cdot 10^{-11}\,\mathrm{m}^2$. The integration limits for the energy integral are set to $E\pm3\sigma_E$ where the $E=1.165\,\mathrm{eV}$ corresponding to 1064\,nm photons and $\sigma_E = 0.088\,\mathrm{eV}$ based on the skewed Gaussian fit to the distribution of $\mathrm{Ph}_\mathrm{FFT}$ for light pulses. With these parameters, the integral in Eq. (\ref{Plancks_Spectrum_Eq}) at $T=295\,\mathrm{K}$ results in $\dot{N}=5.1\,\mathrm{mHz}$. The calculated rate is fivefold higher than what was observed by the CNN from the experimental data. However, the above presented calculation represents the theoretical maximum black-body rate and does not take into account various loss mechanisms present in experimental setup. In reality, this rate is lowered by the limited detection efficiency of the TES together with wavelength dependent transmission losses in the used mating sleeves and the fiber itself. In particular, fiber curling inside the cryostat, that was present in our experimental setup, has been observed to result in significant attenuation towards longer wavelength photons. The simulation of losses due to optical fiber, fiber curling and TES response in the same experimental setup used in this work has been recently addressed in Refs.~\cite{RubieraGimeno2024optimizing, Gimeno2025simulation}. Using the associated simulation pipeline, we estimate a $0.57\,\mathrm{mHz}$ black-body background associated with cut-region $[\mu_\mathrm{m}-3\sigma,\,\mu_\mathrm{m}+3\sigma]$. This corresponds better to the herein estimated value of $1.02\,\mathrm{mHz} \pm 0.2\,\mathrm{mHz}$. Considering the limitations of the simulation, the misclassified dark pulses seem indeed likely to result from near-1064\,nm black-body photons coupled into the optical fiber based on the above presented rate comparison.
\begin{figure}[t!]
\begin{center}
\includegraphics[width=6.0cm]{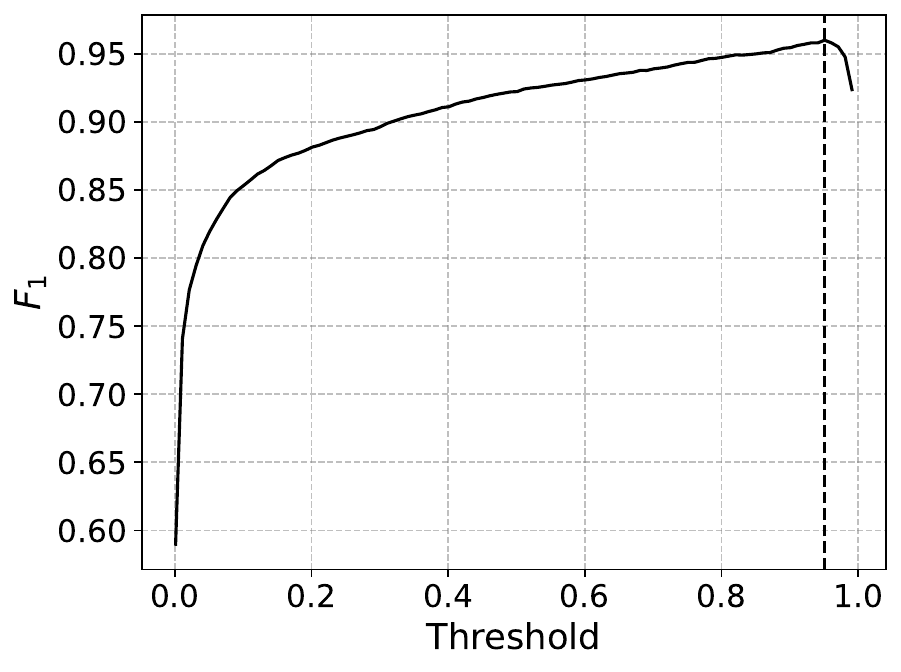}
\caption{(a) The $F_1$ score calculated for the experimental test set (2898 light pulses, 7872 dark pulses) using the best performing model within the CNN~\textit{ensemble~\#2} as a function of threshold. The maximum of $F_1=0.96$ was found for a threshold of 0.95, illustrated by the dashed vertical line in the figure. }
\label{CNN3_F1-threshold_Fig}
\end{center}
\end{figure}

In order to provide more concrete evidence on the origin of the misclassified dark pulses, we investigate them using CNN within the \textit{ensemble \#2} for which the percentage of misclassified dark pulses was closest to the ensemble average $2.00\%$ (157) reported above (see Fig.~\ref{CNN_ensemble_figure_abc}(c)). The corresponding CNN misclassified $2.03\%$ (160) of the 7872 dark pulses under analysis and can be thus considered to represent the average of the ensemble with respect to misclassified dark pulses up to a good extent. It achieves $F_1=0.96$ at an optimal threshold of $0.95$. Finding the optimal threshold is illustrated in Fig.~\ref{CNN3_F1-threshold_Fig}. In this case, the associated curve peaks rather sharply at the optimum $\mathrm{Th.=0.95}$, indicating susceptibility towards the optimization of the threshold.

The average misclassified dark pulse is illustrated in Fig.~\ref{misclassified_dark_pulses_Fig-ab}(a) together with the average light pulse. The shapes of the average pulses closely resemble each other suggesting a source of the same nature. In addition, Fig.~\ref{misclassified_dark_pulses_Fig-ab}(b) illustrates the PCA scatter plot showing the projections of the associated feature vectors ($\tau_\mathrm{rise}$, $\tau_\mathrm{decay}$, $\chi_\mathrm{Ph}^2$, $\mathrm{Ph}_\mathrm{FFT}$, $\chi^2_\mathrm{FFT}$) of the light and misclassified dark pulses onto the two main principal components. The majority of the misclassified dark pulses are clustered in the near vicinity of the light pulses. Only roughly 14 out of the 160 misclassified dark pulses are clearly outside the (red) cluster defined by light pulses. Thus, the vast majority of the misclassified dark pulses are most likely triggered by near-1064\,nm photons originating from the black-body background. I.e., for the CNN they are indistinguishable from the light pulses given the energy resolution of the TES. Having these pulses distort the extrinsics (dark) dataset has detrimental effects on the training of the CNNs as the model is repeatedly being trained to classify an actual light pulse as dark. It is evident that this causes confusion in the training process of the CNN and limits the performance of the model. The presence of near-1064\,nm black-body photon triggers in the dark dataset also explains why the CNNs are so susceptible towards the division of the dataset into training and testing sets (Fig.~\ref{training_and_testing_data_schematics_Fig}). How many of these, technically mislabeled dark pulses, end up in training or testing sets has a significant impact on both training and evaluation of the CNNs. This results in the observed high standard deviations in the evaluation metrics of the CNN \textit{ensemble~\#2} (Table~\ref{ensemble2_evaluation_table}).

In the following section, we will investigate the black-body radiation induced training confusion and show that this ultimately limits the performance of the CNN.

\section{Black-body photons and training confusion}
\label{BB-photons_and_training_confusion_section}
\noindent In the previous section we concluded that the vast majority of the misclassified dark pulses are ultimately near-1064\,nm photons. While these are physically indistinguishable from the light pulses given the energy resolution of the TES, training the CNNs to learn to classify these as dark pulses evidently causes confusion. The detrimental effects of this \textit{label noise} have been widely studied \cite{Liu2015classification, Song2022learning, Im2023binary}. While a small fraction of corrupted labels can result in improved performance by acting as a form of regularization, their effect on learning is generally detrimental \cite{Lee2022binary}. This is particularly true for the herein studied CNNs, which are already regulated by the dropout layer.

In consequence, the reported classifying performances of the CNNs in the previous sections should be considered as lower limits with room for further improvement when trained with an ideal, undistorted dataset. In order to demonstrate this, we relabel the target values of the 160 misclassified dark pulses from the previous section as light (target value $0\rightarrow 1$) and use this data for retraining an ensemble of 10 CNNs exactly as in Section~\ref{background_classification_section} (\textit{ensemble \#2}, see Fig.~\ref{CNN_ensemble_figure_abc}(b)). It should be noted that the 160 misclassified dark pulses form a subset of the testing data used in Section~\ref{background_classification_section} and consequently some near-1064\,nm black-body photon triggers can still be left unidentified in the training set. Moreover, the 160 misclassified dark pulses had roughly 14 pulses which were clearly not clustered in the vicinity of the light pulses and thus these may not correspond to near-1064\,nm black-body photons. Regardless, we relabel all of the 160 previously misclassified dark pulses as light, after which the vast majority of the near-1064\,nm photon triggered dark counts should have been addressed. While the dataset still remains somewhat distorted, one expects to observe improvement in the CNN performance when trained with the relabeled dataset if there was training confusion present previously.

Thus, we proceed in retraining the CNN \textit{ensemble~\#2} (Fig.~\ref{CNN_ensemble_figure_abc}(b)) using the dataset with 160 relabeled dark pulses based on Section~\ref{background_classification_section}. The overall dataset now contains $3898+160=4058$ light pulses and $8872-160=8712$ dark pulses which is then further divided into training and testing data according to Fig.~(\ref{training_and_testing_data_schematics_Fig}). Upon doing this, the $F_1$ score improves from the previously estimated $\langle F_1 \rangle = 0.961 \pm 0.005$ (Table~\ref{ensemble2_evaluation_table}) to $\langle F_1 \rangle=0.974 \pm 0.004$. Evidently, the average $S$ score also improves due to additional background discrimination from $\langle S \rangle = 0.95 \pm 0.23$ to $\langle S \rangle = 1.61 \pm 0.58$ ($\epsilon_\mathrm{a}=85.5\%$ and $n_\mathrm{b}=0.17\,\mathrm{mHz}$), but still does not outperform the cut-based analysis after relabeling the presumed black-body triggers. The observed improvement in the $F_1$ score (see section~\ref{performance_evaluation_section}) is direct evidence that the performance of the CNN is limited by the training confusion caused by the presence of near-1064\,nm black-body photon triggers in the measured extrinsic background. It should be noted that the presence of near-1064\,nm black-body photon triggers in the dark dataset also imposes a physical limit on the classification performance of the cut-based analysis. Analogous to training of the CNNs, the skewed Gaussian distributions determining the cut regions were calculated using 1000 randomly selected light pulses, without any influence from the dark dataset containing the mislabeled dark pulses (see Section~\ref{comparison_to_cut-based_analysis_section}). After fixing the cut regions, the $S$~score was calculated using the rest of the data, including all of the near-1064\,nm black-body photon triggers. Yet still, the cut-based analysis significantly outperforms the CNNs. As we previously concluded that inadequate hyperparameter optimization is not likely to limit the performance of the CNNs (see Section~\ref{hyperparameter_optimization}), this suggests that the CNN’s performance is limited by the training process due to the presence of mislabeled dark pulses.

\section{Conclusions and outlook}
\label{conclusions_section}
\noindent We have aimed to improve the detection significance of a TES by analyzing the experimentally measured 1064\,nm laser photon (light) and extrinsics background event (dark) triggered univariate time traces using a CNN based binary classifier. After hyperparameter optimization, the CNN ensemble resulted in average detection significance of $\langle S \rangle=0.95\pm0.23$, still being outperformed by our previously used (non-ML) cut-based analysis by 36\%. 

We have concluded that inadequate hyperparameter optimization is unlikely to limit the performance of the CNN. Meanwhile, our findings suggest that the limited performance can be attributed to training confusion introduced by the near-1064\,nm black-body photon triggers in the extrinsics background. The CNN's sensitivity to the used training data is particularly manifested as the large standard deviations in the calculated $S$ scores when compared with corresponding values obtained from cut-based analysis. Thus, the used experimental binary dataset seems to be inadequate for training the CNNs for improved noise suppression. Based on our results, we recommend further exploration of regression-based CNNs, with a strong focus on optimizing the size and structure of the training set. This includes, for example, systematically studying how the number and separation of distinct photon wavelengths affect the model’s regression performance when associating a pulse with a given wavelength. Our recently published simulation framework for TES pulses provides particularly great opportunities for the required systematic generation of datasets that correspond well to experimental data \cite{RubieraGimeno2024optimizing, Gimeno2025simulation}. We also want to point out that the use of various ML models in unsupervised fashion, such as neural network based autoencoders, has shown great potential to address background suppression related tasks \cite{Holl2019deep}.

While there exist several potential post-data analysis methods that can improve the detection significance of the TES for the ALPS~II experiment, we argue that reaching the black-body radiation limited \cite{Miller2007superconducting, Gimeno2025simulation} ultra-low background of $10^{-5}\,\mathrm{Hz}$ ultimately requires the implementation of hardware-based background suppression methods. As already suggested in Ref.~\cite{Miller2007superconducting}, the simplest way to do this is to apply a cryogenic narrow bandpass optical filter in front of the TES which effectively improves its energy resolution. We are currently building a cryogenic optical U-bench inside our dilution refrigerator enabling the implementation of such filter as a part of our TES setup.

\section*{Data availibility}
The datasets generated and/or analysed during the current study are available in the \textit{Dataset used for Binary Classification of Light and Dark Time Traces of a Transition Edge Sensor Using Convolutional Neural Networks} repository, https://doi.org/10.5281/zenodo.17347454

\section*{Author contributions}
E.R. implemented the machine learning models, made figures and  wrote the manuscript. J.A.R.G. performed the cut-based analysis. E.R., G.O., J.A.R.G and C.S were involved with acquiring of experimental data. M.M. conceived the project together with K.-S.I,  F.J and A.L, also providing supervision and support through the project. All authors discussed the results, provided feedback and approved the final version of the manuscript.

\section*{Declarations}
\subsection*{Funding}
F.J., A.L. and C.S. acknowledge support from the Deutsche Forschungsgemeinschaft (DFG, German Research Foundation) under Germany’s Excellence Strategy – EXC 2121 “Quantum Universe” – 390833306. M.M. and E.R. acknowledge support from the European Research Council (ERC) under the European Union’s Horizon 2020 research and innovation program, Grant Agreement No. 948689 (AxionDM).

\bibliography{bibliography.bib}
\newpage
\end{document}